# From droop to optimality: The potential of volt/var control for power distribution grid enhancement

Jonas G. Matt [a,*], Lukas Ortmann [b], Saverio Bolognani [a], Florian Dörfler [a]

[a] Automatic Control Laboratory, ETH Zurich, Zürich, Switzerland
[b] Eastern Switzerland University of Applied Sciences, Rapperswil-Jona, Switzerland

## HIGHLIGHTS

- We combine a benchmark distribution grid model with real residential power consumption and rooftop photovoltaics data.
- We use predictions of the increase in rooftop PV capacity to emphasize the future need for more effective volt/var control.
- We demonstrate that local volt/var controllers are inherently suboptimal and will become insufficient in the future.
- We illustrate the near-optimality of coordinated/centralized methods such as Online Feedback Optimization (OFO).
- We show how more advanced local controllers (e.g., optimized based on historical data) may also shrink the gap to optimality.

## ARTICLE INFO

*Keywords:*
Grid enhancement
Volt/var control
Reactive power
Renewable energy
Online feedback optimization
Droop control
Real data

## ABSTRACT

When high amounts of active power are injected into power distribution grids, the overall power flow is limited because voltages reach their upper acceptable limits. Volt/var control aims to raise this power flow limit without physically reinforcing the grid but by controlling the voltage using reactive power. We use real consumption and generation data on a low-voltage CIGRÉ grid model and an experiment on a real distribution grid feeder to analyze how different volt/var methods can enhance the grid. We show that local droop control enhances the grid but underutilizes the reactive power resources. We discuss how this inefficiency can be partly reduced by fine-tuning the droop curves through data-driven techniques but illustrate that inherent trade-off persist for any local control method. We finally demonstrate that coordinated control methods can track the optimal solution and enhance the grid to its full potential if grid-wide communication is available. Our numerical study over a whole year of real data suggests that coordinated volt/var control can enable another 10.4% of maximum active power injections compared to droop control. In a small-scale real-life experiment, coordinated control enhanced the grid by the same amount.

## 1. Introduction

Recent decades have seen a global surge in renewable energy production. The main drivers of this development are solar photovoltaics (PV), wind power and other distributed energy resources (DERs). Historically, the bulk of electric energy production has been centralized in large power plants and transported to the consumers via a one-directional electricity grid. With the surge of DERs, increasing amounts of power are injected into the grid at lower voltage levels. While shifting power consumption in time and storing power generation in local batteries can reduce the peak power flows, some parts of the distribution grids will still reach or have already reached their capacity limit. Today's distribution grids typically consist of electrically long lines, making voltage deviations the main determinant of the grid's capacity. A study found that in Germany, 75% of the necessary grid reinforcements in low voltage distribution grids are due to voltage problems [1, p. 172].

Voltages exceeding their safe limits (overvoltages) not only put at risk the existing infrastructure but may also impede the further deployment of DERs. Grid operators are thus confronted with the need to mitigate them. While constructing additional lines can alleviate the issues, such physical reinforcement of the grid is costly. A more cost-effective alternative lies in the application of smart control strategies, which can enhance the active power flow capacity without the necessity of new lines; an approach we refer to as grid enhancement. In this context, volt/var control emerges as an economical and underutilized option for voltage regulation. Also referred to as reactive power control, it aims at actively absorbing or injecting reactive power into the grid to regulate the voltage. In the past, dedicated devices (e.g., capacitor banks) had to be installed to provide the required reactive power capabilities. However, with the proliferation of DERs also comes an increased capacity of grid-connected power converters, e.g., PV inverters. These devices are controlled voltage sources and can actively engage in volt/var control by adjusting their reactive power setpoints according to a prescribed control scheme.

* Corresponding author.
*Email address:* jmatt@control.ee.ethz.ch (J.G. Matt).

The optimality of a volt/var control scheme can be assessed by comparing its performance to the solution of an optimal reactive power flow problem like

$$\begin{aligned} \textbf{minimize} \quad & \text{reactive power used} \\ \textbf{subject to} \quad & \text{voltage constraints at all buses } \textit{and} \\ & \text{reactive power limits of participating devices,} \end{aligned} \tag{1}$$

which ensures an acceptable voltage profile across the grid whenever available reactive power resources permit it using a minimum amount of reactive power to do so. From a grid enhancement perspective, choosing the reactive power setpoints according to (1) maximizes the potential active power flows without compromising voltage stability.

The predominant method currently employed for volt/var control is droop control, wherein each inverter determines its reactive power setpoint based on a predefined function of the local voltage measurement, commonly referred to as a droop curve. In fact, many modern grid codes already mandate the involvement of these devices in voltage regulation [2–4]. Droop control is a local control scheme because the control actions of participating devices are executed without coordination. Previous work has shown that optimal control according to (1) cannot be achieved under all operating conditions by any local controller [5], that is, without coordination between nodes of the grid.

In practice, this theoretical optimality gap can be reduced (but not eliminated) by appropriate tuning of the local droop curves. Recent advanced solutions propose local droop controllers that are tailored to the specific grid [6], optimized based about the available information on the problem [7–9], or determined by learning the optimal droop curve from ensembles of optimal reactive power flow (ORPF) solutions [10–12]. We refer to [13] for a comparative review of these and other similar methods. Typically, these data-driven approaches perform well in the scenarios for which they have been trained/optimized but are farther from optimality outside those operating conditions.

An alternative avenue for improving volt/var control involves coordinated control strategies, thanks to advances in the real-time communication infrastructure. It has been shown that, when communication is available, it is possible to design centralized controllers that steer the grid to the solution of the ORPF problem. We denote this broad class of control schemes as Online Feedback Optimization (OFO) strategies and we refer to [14] for a review of these methods. Generally, OFO volt/var control strategies repeatedly update the reactive power setpoints of the DERs based on real-time voltage measurements and they do not require a detailed grid model or disturbance information. They come with theoretical convergence guarantees, showing that they steer the grid to the solution of (1) and track it even in the presence of time-varying disturbances. They have been tested experimentally on power system testbeds [15–21] and an instance is running 24/7 on a real distribution feeder [22].

Despite the availability of various advanced control schemes, several questions remain unaddressed. To our knowledge, the interplay of the simultaneous increase in active power infeed and control authority, caused by the increase in distributed energy resource (DER) capacity, has not yet been sufficiently analyzed. Furthermore, while the theoretical limitations of local control schemes have been proven, it remains unclear whether the optimality unlocked by coordination is worth investing in real-time communication infrastructure. Therefore, we see a need for a realistic and direct comparison of these different paradigms for volt/var control.

In this paper, we thus aim to investigate the optimality gap of different volt/var control schemes in a realistic setting. Using an extensive dataset of real household electricity consumption and PV production, combined with predictions of increasing PV capacity over the next two decades, we first demonstrate that overvoltages will occur more frequently during peaks of DER infeed. We then show that even though the rise of DERs comes with increased control capacity (reactive power), the currently deployed volt/var methods will not suffice to manage the corresponding increase in active power. This highlights the importance of more effective volt/var control strategies in the future. We investigate two possible solutions: improving on the local control methods stipulated by current grid codes or coordinating the control actions of controllable devices to achieve near-optimal control. We implement the methods proposed by Gupta et al. [12] and Yuan et al. [23] as two recent examples of how the optimality gap of local control can be reduced even in the absence of communication. Then, we demonstrate that when a communication channel is available, centralized control methods like OFO can accurately track the ORPF solution, resulting in optimal grid enhancement.

The remainder of this paper is structured as follows: In Section 2, we describe the setup of the numerical study and define the studied volt/var controllers in Section 3. We present the results of the numerical study in Section 4 and the experimental results in Section 6. We discuss the fundamental trade-off and method comparison in Section 5, before concluding the paper in Section 7.

## 2. Simulation setup

The simulation framework is based on Python and the open-source package `pandapower` [24]. All code is available in a public repository.[1]

### 2.1. Benchmark low voltage grid

At the foundation of the simulation framework lies the benchmark low voltage distribution grid with a European layout, as proposed by CIGRÉ [25]. The grid has a radial structure, which is typical for European distribution grids. As shown in Fig. 1, only the residential subnetwork is used, which is characterized by underground cable transmission, a line-to-line voltage of 400 V, and a system frequency of 50 Hz. A transformer connects the network to the external MV grid, modeled as a constant voltage source at 1 p.u.

The grid layout comprises 18 measurement buses, five of which are the connection points of loads and local PV generation. Each for these five load buses aggregates a small neighborhood of buildings with peak load values ranging from 15 to 55 kW per bus.

### 2.2. Data

To quantify how the different volt/var methods perform, real data were used for both household electricity consumption and PV generation. Among the limited range of publicly available datasets suited for this purpose, Dataport by Pecan Street Inc. [26] was identified as the best one according to the following criteria: a large number of different households at a single location with matching load and PV generation data, high temporal resolution and a total duration sufficient to reveal seasonal effects.

The dataset available under free academic licensing is from the year 2018, and contains data from 25 different households in Austin, USA with a temporal resolution of one second and a total duration of one year. It was downsampled to a resolution of one minute to address the high cost of computing the power flows at each time step. The households available in Dataport were assigned to the five load buses of the benchmark grid such that the aggregated peak demand and generation values were in accordance with the peak values stated by CIGRÉ. Fig. 2 shows the resulting total load and generation at each bus for one exemplary day.

[1] Code repository: gitlab.ethz.ch/jmatt/volt-var.

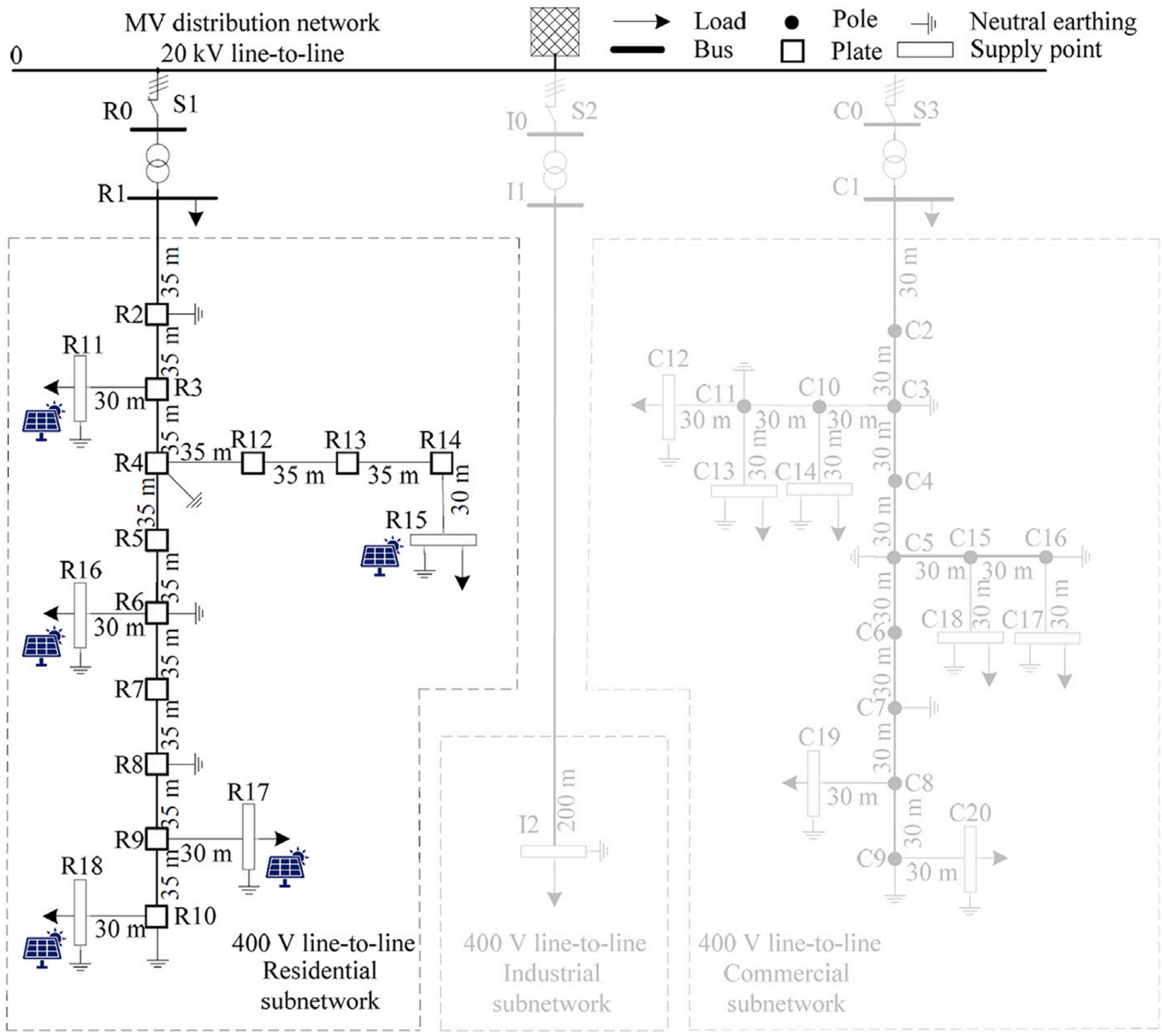

**Fig. 1.** The CIGRÉ low-voltage distribution grid with a European layout. The highlighted residential subnetwork was used for this work. The dark blue icons indicate the locations of loads and PV power infeed. The figure has been adapted from [25]. (For interpretation of the references to colour in this figure legend, the reader is referred to the web version of this article.)

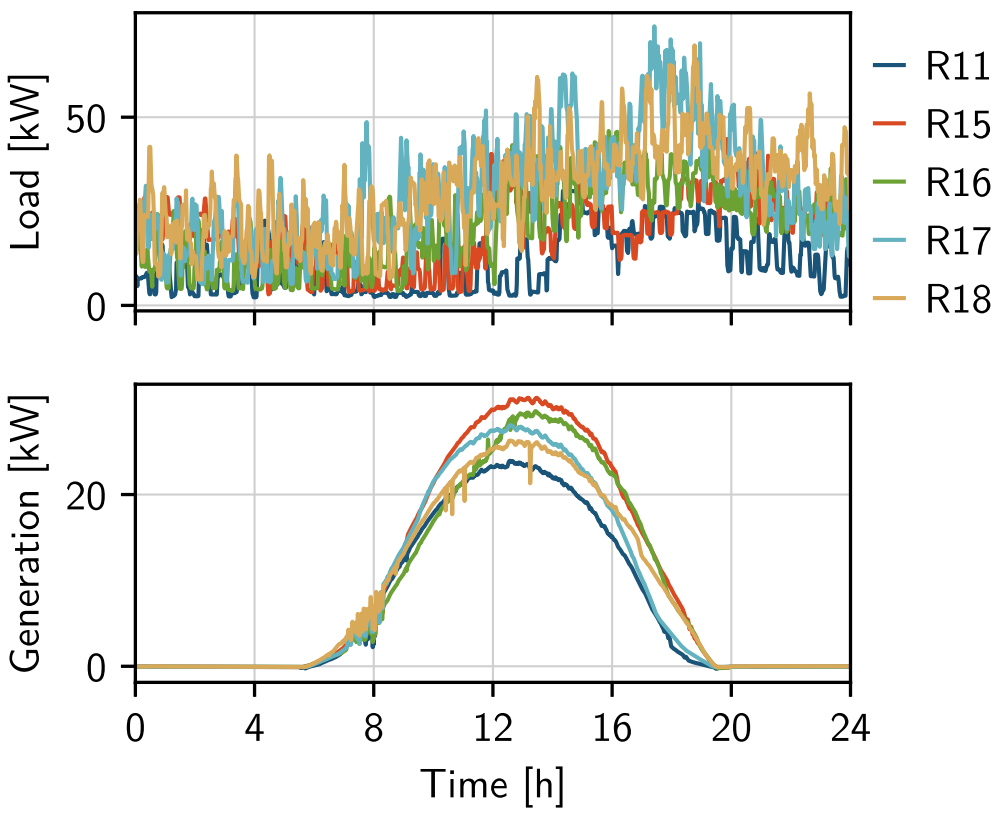

**Fig. 2.** One exemplary day (July 2, 2018) from the resulting dataset, after assigning the available Dataport data to the load buses in the CIGRÉ LV grid. All other buses have neither load nor generation.

### 2.3. Scenarios for PV integration

To investigate the impact of increasing power injections from DERs into distribution grids, three scenarios have been designed which differ in the amount of installed PV capacity. Starting from the base scenario given by the actual PV data from the year 2018, PV capacity (both active and reactive power) is increased by the factors 2 and 3.5, to create the PV integration scenarios 2030 and 2035, respectively. These values are in accordance with current average predictions for the global increase in PV installations [27–30].

### 2.4. Modeling assumptions

The numerical results presented in this work are subject to several simplifying assumptions, made deliberately to keep the study tractable and conceptually simple while ensuring that the relative performance of the compared methods can still be evaluated in a realistic way.

*Voltage constraints only.* Our analysis focuses exclusively on voltage constraint violations as the limiting factor for PV integration. Line flow limits are not considered, which is justified for the low-voltage network studied here, where overvoltage becomes the binding constraint before thermal limits are reached.

*Reactive power capability region.* We assume a box-shaped P-Q feasibility region, so that reactive power setpoints can take any value in $[q_{\min}, q_{\max}]$ regardless of active power generation. More realistic capability curves would shrink the reactive power limits at high active power output. However, since all methods draw on the same physical resources, this affects them similarly and the relative ranking is not expected to change.

*Network size and phase balance.* The CIGRÉ benchmark network represents a moderately sized low-voltage feeder. While larger or more complex network topologies could in principle affect the quantitative results, the used benchmark is rich enough to compare the behavior of different control strategies. Phase coupling is neglected and the network is treated as a balanced single-phase equivalent. While unbalance effects are certainly present in real low-voltage grids, this simplification is standard practice in the volt/var control literature and permits a tractable analysis over a full year of high-resolution data without affecting the relative comparison between control strategies.

**Table 1**
Persisting constraint violations in per-unit hours under the droop and OFO controllers, for uniform PV scaling and for the worst-case distribution concentrating all additional PV capacity at the electrically most isolated bus R15.

| Scenario | Controller | Uniform | Worst-case |
|---|---|---|---|
| 2030 | Droop | 4.1 | 228.4 |
| | OFO | 3.9 | 168.8 |
| 2035 | Droop | 32.3 | 524.5 |
| | OFO | 23.7 | 409.3 |

*Uniform scaling of PV capacity and no load scaling.* The 2030 and 2035 scenarios scale PV capacity uniformly across all buses. In practice, new installations could be distributed unevenly. A more uneven increase in PV capacity further widens the gap between coordinated controllers (such as OFO) and local controllers (such as droop, the default local controller, which we present in detail in Section 3.2). Table 1 illustrates this for the worst-case distribution, in which all additional PV capacity is installed at R15, the most electrically isolated bus in the grid. That is, the voltage at R15 is sensitive to local active and reactive power injections but not much so to injections at any other bus. Under this distribution, the difference in persisting constraint violations between the droop and OFO controllers increases from 0.2 to 59.6 per-unit hours in the 2030 scenario and from 8.6 to 115.2 per-unit hours in the 2035 scenario, compared to the default uniform scaling. The performance of the other, optimized droop controllers must lie in between: being local controllers, they remain suboptimal relative to OFO, but they should not perform worse than standard droop, whose tuning they can recover as a special case. As the relative order of the methods in terms of performance is therefore not affected, we don't examine alternative spatial distributions.

Furthermore, we hold the load profiles fixed across scenarios. In reality, the electrification of heating and transport will increase consumption, partially offsetting net PV injections. Our fixed-load assumption therefore represents an upper bound on the severity of overvoltages and the associated need for volt/var control. However, given that we use only average PV growth projections and that load and generation increases will only partially coincide in time, this upper bound is reasonably tight.

*Either no or perfect communication.* We deliberately study two polar cases: local controllers that operate without any communication, and coordinated controllers (OFO, ORPF) with perfect (lossless, delay-free) communication to all grid nodes. This choice is motivated by the goal of quantifying the fundamental performance gap between local and optimal control, which can only be fully closed with complete grid-wide information. The spectrum in between, such as distributed controllers or imperfect communication, is of practical interest but lies outside the scope of this work.

## 3. Reviewed methods

First, we define a benchmark that represents the theoretical limit of what can be achieved using only reactive power resources for voltage control. This corresponds to the solution of an ORPF and can be computed based on the exact knowledge of the grid data.

We then compare the performance of four different volt/var controllers within the proposed simulation framework: (1) local droop control with standard parameter tuning, (2) a local droop controller tuned based on historical data and a system model according to [23], (3) a local droop controller tuned based on historical solutions of the ORPF problem according to [11], and (4) OFO, which relies on full communication and measurement feedback to track the solution of the ORPF problem.

### 3.1. Benchmark: optimal reactive power flow

In the simulation framework, a flawless model of the grid is available, and the optimal reactive power compensation can be computed via an ORPF. Its solution is the vector of reactive power setpoints $q$ that solves the constrained optimization problem

$$\begin{aligned} &\underset{q}{\operatorname{argmin}}\ \|q\|_2^2 \\ &\quad \text{s.t. } v_{\min} \le v_h(q,w) \le v_{\max} \quad \text{at all buses } h \\ &\qquad\quad q_{h,\min} \le q_h \le q_{h,\max} \qquad \text{at all buses } h, \end{aligned} \tag{2}$$

where $v_h(q,w)$ is the bus voltage at bus $h$ for a given reactive power injection $q$ by all inverters and a certain disturbance $w$. The voltage constraints as defined by the grid code are given by $v_{\min}$ and $v_{\max}$ and are assumed to be the same at each bus. Finally, the aggregated lower and upper reactive power limits of all inverters connected to bus $h$ are denoted as $q_{h,\min}$ and $q_{h,\max}$, respectively. Overall, the ORPF approach attempts to keep all voltages within the constraints while using a minimum amount of squared reactive power. If the combined reactive power capabilities of all PV inverters are insufficient to achieve constraint satisfaction, the ORPF problem becomes infeasible. In that case, the voltage constraints are turned into soft constraints by means of a penalty function. As a result, the reactive power resources operate at the reactive power limit, and the voltages are kept as close to the constraints as possible.

It is important to remark that this is only intended as a theoretical benchmark. Solving the ORPF problem repeatedly in real time necessitates the availability of a precise grid model (topology, line parameters) and accurate real-time disturbance measurements (active and reactive power injections) from all grid nodes. In practice, neither is available. We refer to Section 6 for experimental evidence of this. Furthermore, solving an ORPF problem is computationally expensive, which can render it infeasible in an online setting.

### 3.2. Standard droop control

Droop control is a purely local control scheme in which every PV inverter is required to follow a predefined droop curve that maps the bus voltage at its own point of connection to a reactive power setpoint. Usually, these droop curves take the form of a piecewise linear function such as the one shown below. It is characterized by a deadband around the nominal voltage and maximum injection/consumption of reactive power whenever the voltage constraints are violated. In control language, droop control behaves like a nonlinear P-controller, whose gain is determined by the slope of the droop curve.

$$q_h = \begin{cases} q_{h,\max} & v_h < v_1 \\ q_{h,\max}\dfrac{v_2 - v_h}{v_2 - v_1} & v_1 \le v_h \le v_2 \\ 0 & v_2 \le v_h \le v_3 \\ q_{h,\min}\dfrac{v_h - v_3}{v_4 - v_3} & v_3 \le v_h \le v_4 \\ q_{h,\min} & v_4 < v_h. \end{cases}$$

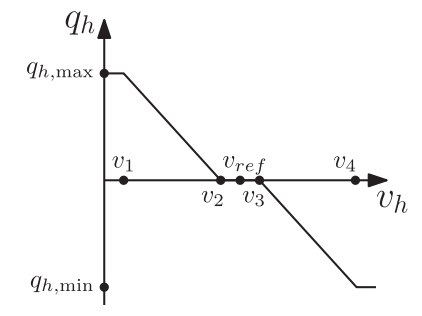

Here, $q_h$ and $v_h$ denote the reactive power injection and voltage at bus $h$, and $q_{h,\min}$ and $q_{h,\max}$ are the total reactive power limits of the inverters at bus $h$, respectively. As the standard tuning, we use $v_1 = v_{\min} = 0.95$, $v_2 = 0.99$, $v_3 = 1.01$, and $v_4 = v_{\max} = 1.05$ (all in p.u.).

### 3.3. Optimized droop control

A natural way to improve droop control performance is to adapt the droop curves, potentially for each inverter individually. Various methods have been proposed to achieve this by relying on grid models, historical data, or a combination of both. In this work, we focus on two representative recent approaches: (1) the method proposed by Gupta et al. [12], which trains a deep neural network (DNN) to determine the optimal inflection points of a standard droop curve using historical generation and load data together with a linearized grid model; and (2) the method of [23], which trains a multi-layer perceptron (MLP) to approximate the optimal reactive power flow (ORPF) solution by fitting droop curves directly.

To enable the employment of steeper droop curves while preserving closed-loop stability, the optimized curves are employed as incremental droop controllers. The reactive power setpoints are updated as a convex combination of the previous setpoints and those prescribed by the optimized droop curves:

$$q_h(t) = (1-\beta) q_h(t-1) + \beta q_{h,\mathrm{droop}}(v_h(t)), \tag{3}$$

with $\beta = 0.8$. The value of $\beta$ lies below the theoretical stability limit given in [31], which depends on the grid sensitivity matrix and the maximum absolute slope of the optimized droop curves, but is large enough to reject disturbances after only a few time steps.

### 3.3.1. DNN droop control

First, we implement a DNN-based method for optimizing droop curves, proposed by Gupta et al. [11]. This method takes as input a linearized grid model along with historical load and generation data. A DNN is then trained to identify the optimal inflection points of the droop curves for the inverters at each bus. The piecewise-linear activation functions of the DNN directly correspond to the piecewise-linear structure of the droop curves. The DNN is only used offline to derive the optimized droop curves, which are subsequently applied in online operation as standard incremental droop controllers.

For training, we use historical load and generation data from every other day in the Dataport dataset, yielding 183 days at 1-minute resolution, or 263′520 data tuples $(p_h, q_h) \in \mathcal{D}_h$ at each bus $h$ in total. A linearized grid model $v(q, p)$, representing voltage sensitivities with respect to active and reactive power at all buses, is used to predict voltage responses to reactive power setpoints. A separate DNN is trained for each bus to determine the optimal inflection points $\theta_h = (v_1^h, \ldots, v_4^h)$ for the local droop curve at bus $h$.

The DNN training objective is to minimize a cost function of the form

$$\arg\min_{\theta_h} \sum_{(p_h, q_h) \in D} \|q_h(v_h; \theta_h)\|_2^2 + \lambda p(v_h), \tag{4}$$

where the penalty function

$$p(v_h) := \begin{cases} (v_{\min} - v_h)^2, & \text{if } v_h < v_{\min}, \\ (v_h - v_{\max})^2, & \text{if } v_h > v_{\max}, \\ 0, & \text{otherwise} \end{cases} \tag{5}$$

acts as a soft approximation of the voltage constraints in (2). The weight $\lambda$ regulates the trade-off between minimizing reactive power use and reducing voltage constraint violations. While the formal optimization problem imposes constraints hierarchically, that is, strictly enforces them whenever feasible, this corresponds to $\lambda \to \infty$, which is computationally intractable. We adopt $\lambda = 300$ as a default, and provide Pareto curves for alternative choices of $\lambda$ in Section 5.

The procedure is repeated for each PV integration scenario separately, where the scaled PV data is incorporated into the training process. Fig. 3 depicts the resulting droop curves for the 2035 scenario.

### 3.4. MLP droop control

An alternative approach to droop curve optimization relies on solving the ORPF problem for historical operating points and subsequently fitting local relationships between voltage and reactive power to the ORPF solutions. Specifically, for each bus $h$, the optimal reactive power setpoints obtained from the ORPF are approximated by a function $q_h(v_h; \theta_h)$, typically via least-squares fitting. While [10] proposed a piecewise-linear fitting procedure, we adopt the method of [23], which employs a multilayer perceptron (MLP) to perform this function approximation.

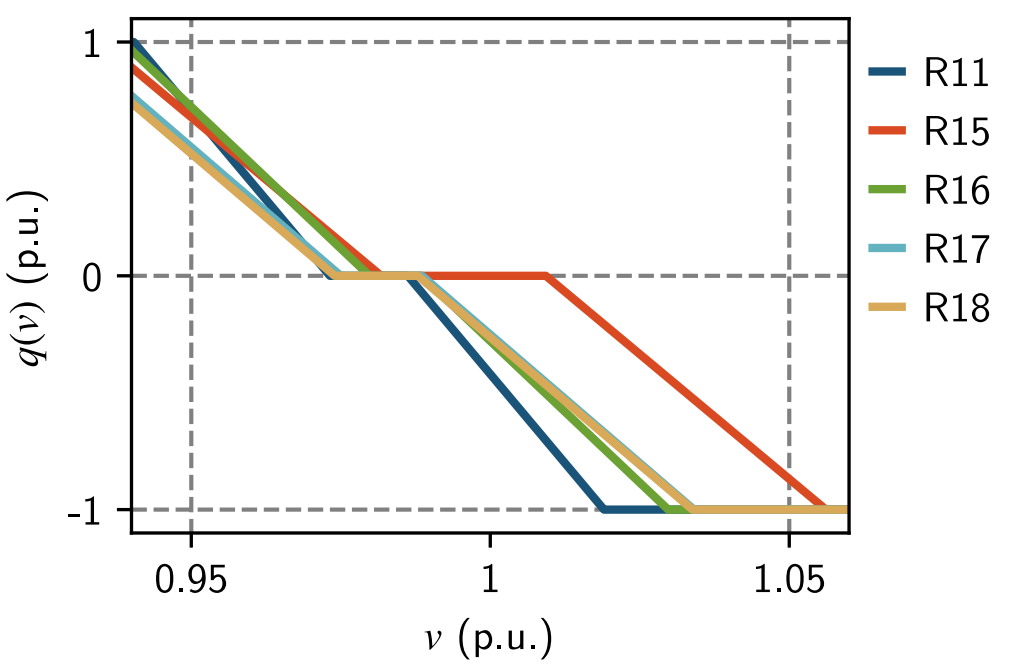

**Fig. 3.** DNN-derived droop curves for the 2035 scenario with $\lambda = 300$.

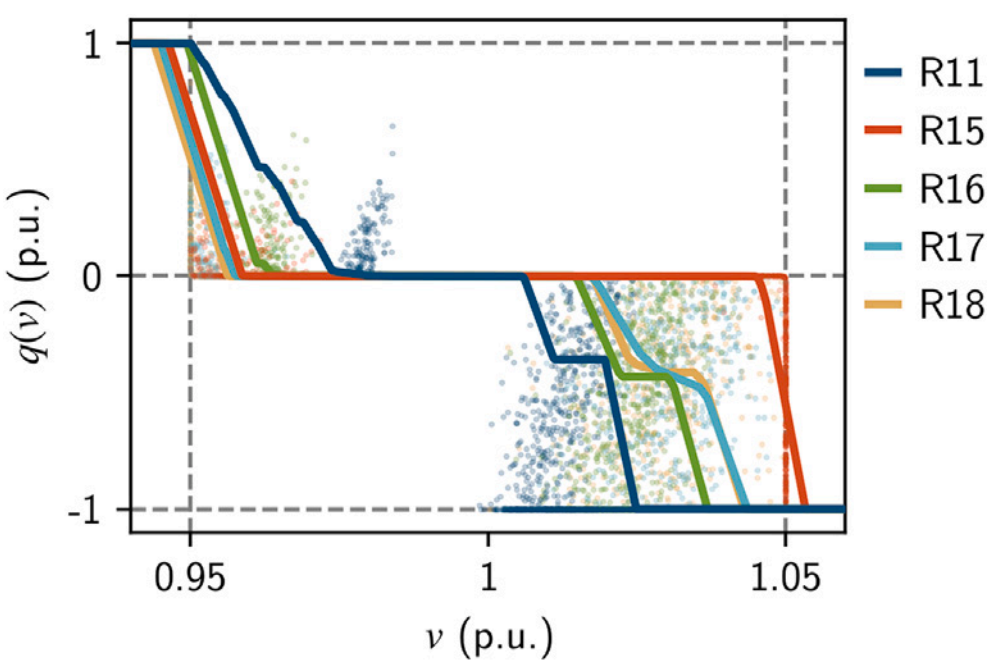

**Fig. 4.** MLP curves for the 2035 scenario.

Training is carried out using the same dataset as for the DNN method: every other day from the available year of Dataport data. The training targets are the reactive power setpoints obtained from the ORPF benchmark described in Section 3.1.

The training procedure is again repeated for each bus and each PV integration scenario, separately. Fig. 4 shows the droop curves produced by the MLP for the 2035 scenario, fitted to the ORPF-derived reactive power setpoints.

### 3.5. Online feedback optimization

OFO is a coordinated control method that can steer a system to the solution of a constrained optimization problem. In the presented case, the ORPF problem (2) is used as the steady-state specification for the OFO design. However, the resulting feedback law does not require a detailed model of the system, as the controller does not need to explicitly solve the ORPF problem. Instead, OFO exploits voltage measurements to gradually update the inputs toward the optimum; the computation of a suitable step requires substantially less model information and no measurement of the loads of the grid, as explained hereafter.

In this paper, we use the OFO controller from [17], which employs a dual ascent strategy to deal with constraints. We assume full observability of the voltage magnitudes at all buses. Partial observability is possible, but asymptotic constraint satisfaction is only guaranteed at locations where the voltage magnitude is measured.

A block-diagram of the method is shown in Fig. 5. Two dual variables per inverter, $\lambda_{\min}$ and $\lambda_{\max}$, integrate constraint violations at each time step according to

$$\lambda_{\min}(t+1) = [\lambda_{\min}(t) + \alpha(v_{\min} - v(t))]_{\geq 0} \tag{6}$$

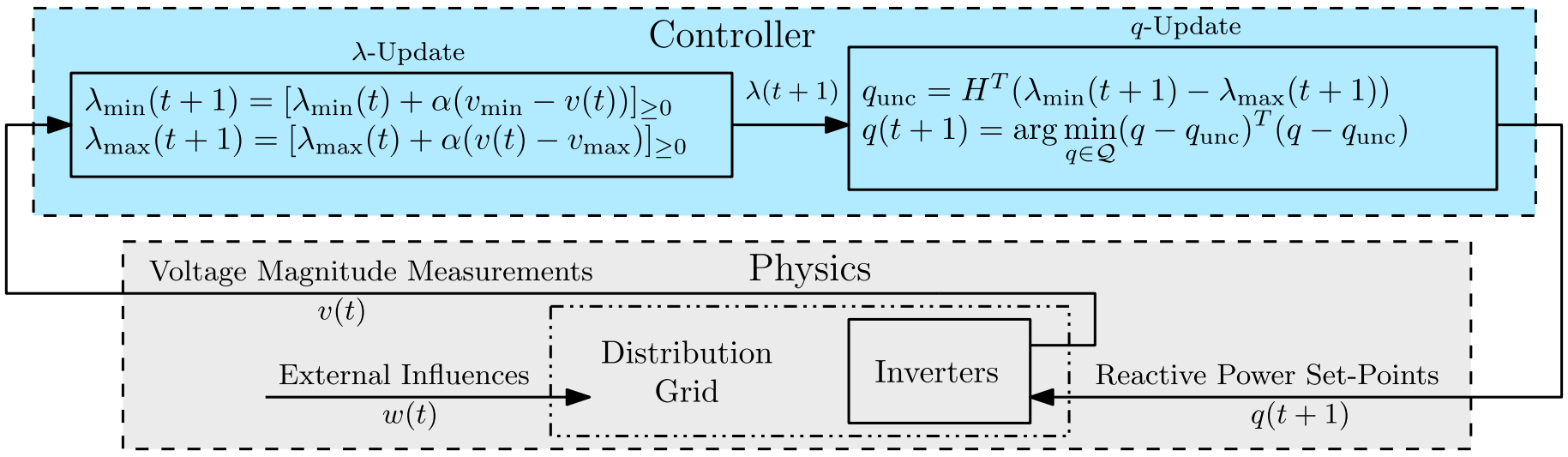

**Fig. 5.** Block diagram of the OFO controller. Two dual variables integrate constraint violations and are used to update the control inputs toward the optimum of the ORPF problem based on closed-loop measurements. The sensitivity matrix $H$ is the only required model information. Figure taken from [17].

$$\lambda_{\max}(t+1) = [\lambda_{\max}(t) + \alpha(v(t) - v_{\max})]_{\geq 0}. \tag{7}$$

The new reactive power setpoints are then determined based on the current values of the dual variables and a sensitivity matrix $H$ as follows:

$$\begin{aligned} q_{\text{unc}} &= H^T(\lambda_{\min}(t+1) - \lambda_{\max}(t+1)) \\ q(t+1) &= \arg\min_{q\in\mathcal{Q}}\ (q - q_{\text{unc}})^T(q - q_{\text{unc}}), \end{aligned} \tag{8}$$

where $\mathcal{Q} = \{q \mid q_{\min} \leq q \leq q_{\max}\}$. More specifically, the matrix $H$ captures how the bus voltages change for a change in reactive power at each bus. It represents a linearization of the power flow equations and can be computed analytically based on an approximate grid model [32], determined by experimental system identification, or learned online [33–35]. In practice, OFO controllers are typically robust against an inaccurate choice of $H$. It may be determined for a nominal and known operating point such as the one without any load and generation as was done in the presented case. However, it has been shown that OFO performs well even ignoring any prior information on the grid topology (that is, setting $H$ equal to the identity matrix) [17].

OFO is further characterized by the step size $\alpha$ which determines the size of the dual ascent update and functions as an integral control gain. Larger values of $\alpha$ increase the convergence speed whereas excessively large values can render the system unstable. For theoretical stability bounds on $\alpha$, we refer the interested reader to [14]. Overall, this yields a trade-off and tuning the step size is hence required to obtain good performance. For the presented simulations, $\alpha = 4$ has been chosen. Finally, we note that OFO is computationally cheap and can be run at small time intervals and on low-power microcontrollers.

## 4. Numerical study of grid enhancement

First, the characteristic behavior of the different control methods is illustrated based on intraday simulations of a particular summer day in the 2035 scenario. Subsequently, the performance of the controllers is compared across the three PV integration scenarios, pointing out the importance of improving the optimality of volt/var control. Finally, a different perspective on the matter is presented: we quantify by how much the PV generation fed into distribution grids can be increased by applying each of the reviewed volt/var control methods. This is to reveal the economic impact that optimal volt/var control can have for grid operators.

### *4.1. Controller behaviors*

In 2035, the active PV power fed into the grid is expected to reach levels at which overvoltages occur frequently, despite the larger availability of reactive power. In fact, none of the proposed volt/var controllers (including the ORPF baseline) is able to keep the voltages within bounds at all times. However, the amount of constraint violations varies drastically for different controllers. This makes this scenario well-suited to observe how the different methods cope with voltages approaching and exceeding the constraints. At the same time, this shows the future need for better volt/var control methods.

Fig. 6 shows the intraday evolution of the voltage and reactive power injection at each bus. Negative values of the latter correspond to the absorption of reactive power by the PV inverters. The amounts of reactive power that are injected or absorbed at each bus are commanded by the ORPF benchmark (Fig. 6(f)), standard droop control (Fig. 6(b)), deep neural network (DNN)-tuned droop control (Fig. 6(c)), multi-layer perceptron (MLP)-tuned droop control (Fig. 6(d)), and OFO (Fig. 6(e)), respectively. The irradiation and active power consumption data are from July 2, 2018 and represent a typical warm and sunny day. This is the data shown in Fig. 2, except that PV generation is scaled by 3.5 in the 2035 scenario. Around noon, the irradiance peak leads to high PV active power production, causing overvoltages to occur across the grid. In the evening, high power consumption and low irradiance give rise to marginal undervoltages. In the following, we will focus on the controller behavior with respect to overvoltages.

Fig. 6(f) shows the results for the ideal ORPF benchmark, which corresponds to the best performance achievable by any volt/var controller under perfect knowledge of the grid model and all active and reactive consumption and generation. When no constraints are violated, no reactive power is used. Once the voltages exceed the constraints, the ORPF solution corresponds to the minimum amount of reactive power that is required to keep them right at the constraints. Once the optimization problem becomes infeasible, i.e., voltage constraint violations become unavoidable, the full reactive power resources are used to minimize the violation of the voltage constraints.

Standard droop control (Fig. 6(b)) applies control inputs based solely on the local bus voltage. Maximum reactive power is used only if the *local* voltage constraint at a bus is violated. However, the inverters that are connected to a bus that is close to the external grid never experience such high voltages (see, for example, bus R11). Hence, they do not absorb much reactive power even though there might be overvoltages occurring further into the grid. Thus, constraint violations persist even if problem (2) is feasible, meaning the voltage violation could be mitigated using other reactive power sources differently. Additionally, some reactive power is used unnecessarily at times when all voltages are within bounds and would have been admissible using lower or even no control effort (for example before 8:00 or after 18:00).

The MLP-optimized droop controller (Fig. 6(d)) improves upon standard droop mainly in terms of efficiency. It uses significantly less reactive power when no voltage constraints are at risk of being violated. However, the number of constraint violations is about the same as with standard droop. In contrast, for the default tuning of $\lambda = 300$, the DNN-optimized droop controllers (Fig. 6(c)) use the reactive power resources more aggressively, resulting in larger reactive power usage but fewer constraint violations than standard droop control. Moreover, the balance between the different nodes is visibly improved, if only to a limited extent. For instance, the reactive power resources at R11, the bus closest to the external grid, are used more actively than with standard droop control.

Lastly, the OFO controller (Fig. 6(e)) exhibits a performance that is comparable to that of the ORPF benchmark. Due to being an integral-like controller, the closed-loop convergence of OFO is not immediate, as constraint violations must be integrated for at least one time-step to initiate an update of the set points. Hence, fast changes in the grid can cause temporary constraint violations until OFO has converged and then satisfies all constraints.

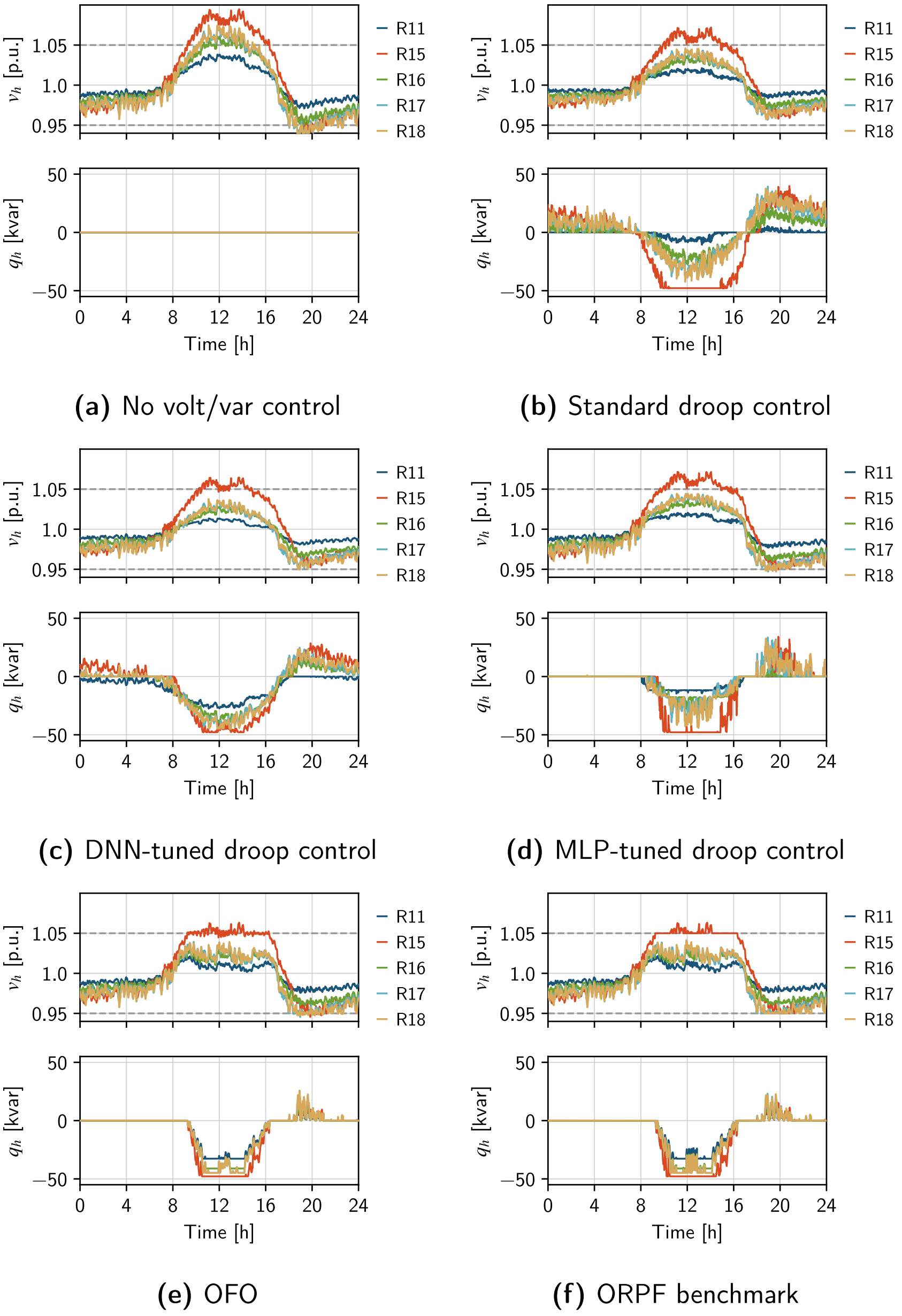

**Fig. 6.** Voltages and reactive power injections throughout a sunny summer day for the 2035 scenario for each of the considered controllers.

### *4.2. Long-term controller performance*

This section analyzes the performance of the controllers throughout a full year. For this purpose, all the available data are used, corresponding to one year of household power consumption and PV power production. This is to reveal the cumulative effect that different volt/var control strategies have on grid operation. Fig. 7 summarizes the results. In the 2020 scenario, corresponding to the PV capacity that is present in distribution grids today, no significant overvoltages occur over the entire year. This indicates why local droop control, currently the state-of-the-art volt/var control method, has been sufficient until now. However, as PV capacity increases in the 2030 and 2035 scenarios, the size and duration of constraint violations increase drastically. Even though the reactive power capacities and hence the control capabilities increase by the same factor as active power production, they no longer suffice to mitigate the overvoltages (even in the ORPF benchmark). Overall, this shows the importance of moving toward better volt/var control methods in the future that use the available reactive power resources as effectively as possible.

Standard droop control yields the worst performance in all three scenarios, allowing the most constraint violations while using the largest amount of reactive energy over the yearly period. The DNN-tuned droop controllers mitigate constraint violations well but require large amounts of reactive energy to do so. While this trade-off can be shifted by tuning the penalty parameter $\lambda$, the resulting Pareto curve lies well below the ORPF solution (see Section 5). MLP-tuned droop control achieves more efficient control with respect to the reactive energy used but allows similar amounts of constraint violations as standard droop control.

The OFO controller outperforms the local control schemes significantly in terms of both better constraint satisfaction and higher

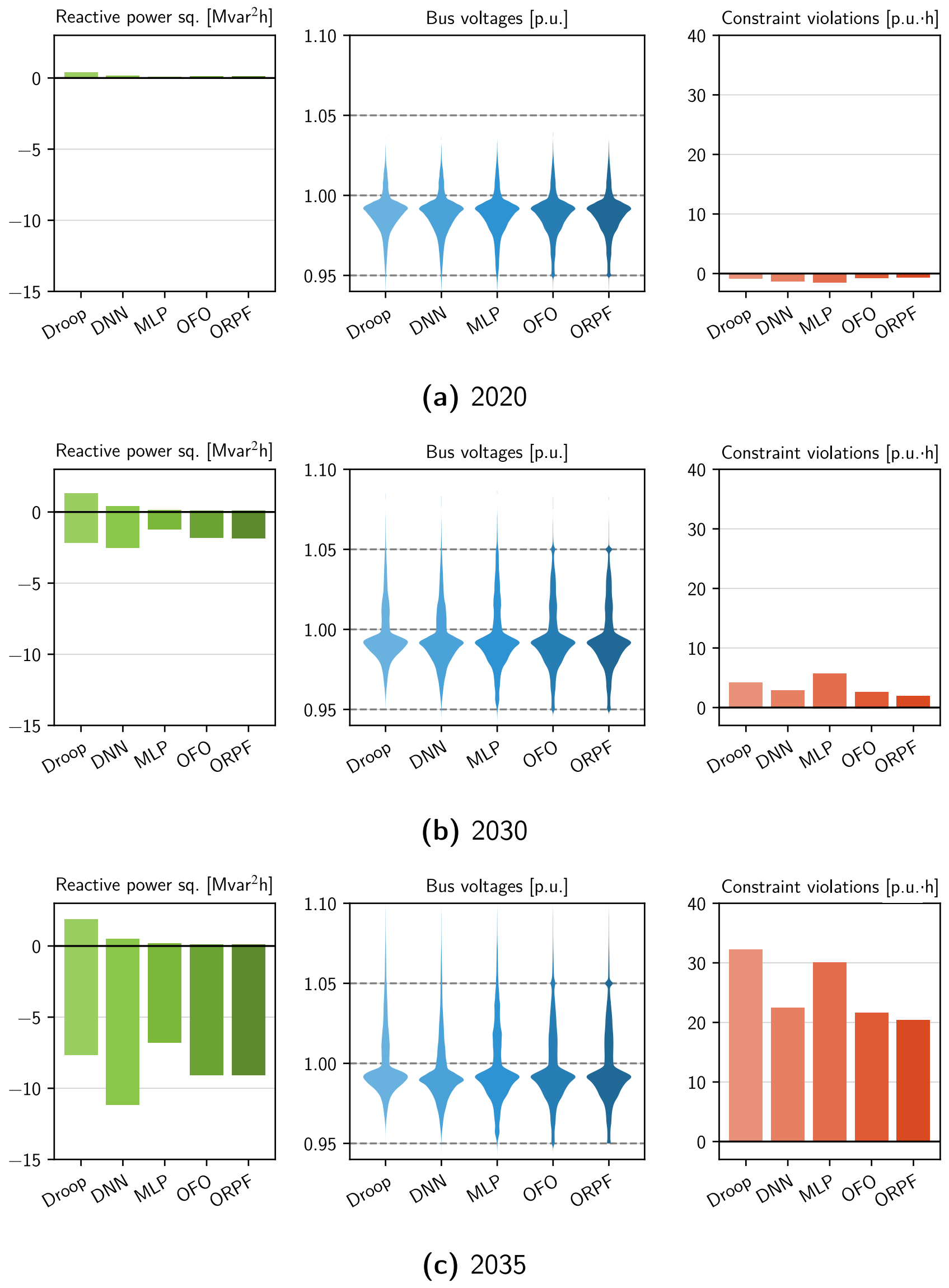

**Fig. 7.** Visualization of the yearly reactive power usage, voltage distributions, and integrated constraint violations for droop, DNN- and MLP-tuned droop, OFO, and ORPF, separately for each of the three scenarios. Positive and negative constraint violation values represent overvoltages and undervoltages, respectively. For the reactive energy use, the reactive power values are squared at each time instance and then summed. This corresponds directly to the cost of the ORPF problem. Positive and negative bars represent reactive energy injected and absorbed by the PV inverters, respectively.

efficiency. Overall, it achieves near-optimal control relative to the ORPF benchmark.

## 5. Discussion of fundamental trade-offs

Any volt/var control method must fundamentally navigate a trade-off between reactive power usage and voltage constraint satisfaction. The ORPF benchmark defines the optimum: it identifies the reactive power setpoints that minimize constraint violations for the least possible reactive power expenditure. The central question is how closely each of the studied methods can approach this optimum, and what prevents them from doing so. Fig. 8 depicts the optimality gap for the different methods.

The ORPF is a network-level problem whose solution depends on the state of all buses simultaneously. A centralized controller such as OFO which has access to grid-wide measurements can in principle track this solution, as can be seen in Fig. 8. Local controllers however, act only on locally available voltage measurements, constituting an information asymmetry that is the fundamental reason for their suboptimal performance. A local controller must rely on a fixed mapping from local voltage to reactive power output that must perform well across many operating conditions. While tuning of the droop curves can improve the average performance, no single choice of curves can be simultaneously optimal for all operating conditions.

The MLP controller attempts to reduce the gap to the ORPF by training droop curves to approximate the ORPF solution itself. The training minimizes average approximation error, which, in particular, does not translate into good constraint satisfaction. For example, the best-fit curves from Fig. 4 do not saturate at $q_{\min}$ or $q_{\max}$ when voltages approach their limits.

The DNN tuning approach optimizes the trade-off directly via a regularization parameter $\lambda$ that balances reactive power usage against constraint satisfaction. Varying $\lambda$ yields the Pareto curves in Fig. 8. As $\lambda$ increases, constraint satisfaction improves toward the ORPF benchmark, but for any fixed reactive power budget, it remains suboptimal.

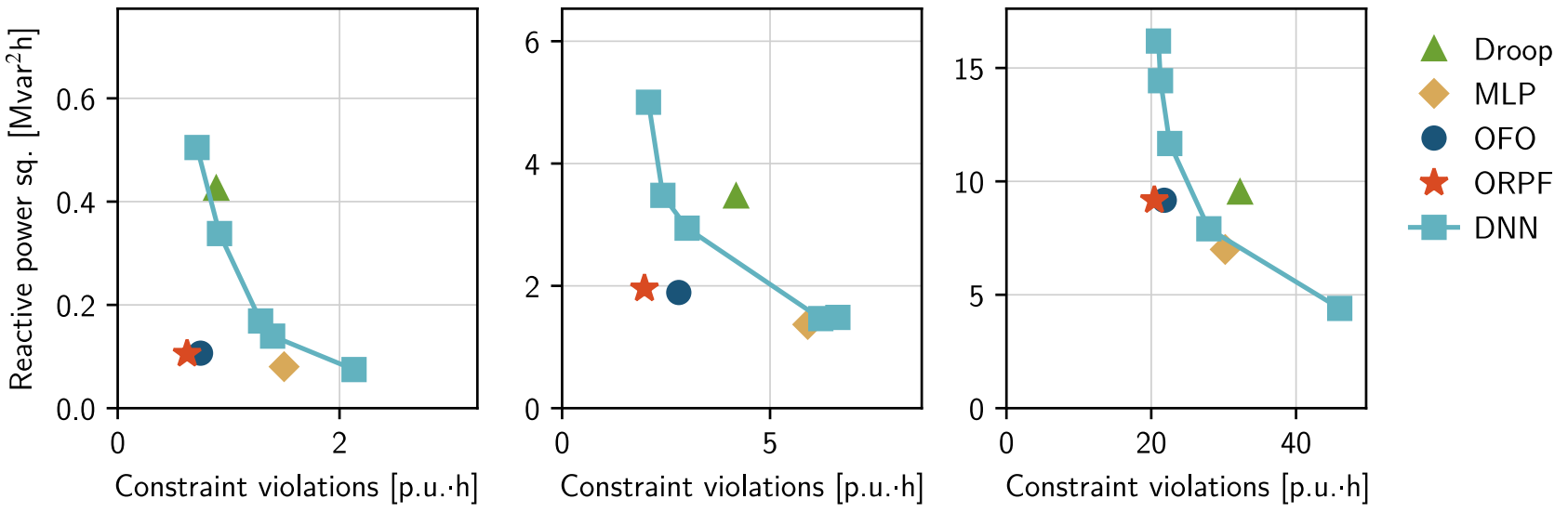

(a) Deployed in 2020 (b) Deployed in 2030 (c) Deployed in 2035

**Fig. 8.** Ability of the studied methods to navigate the trade-off between voltage constraint satisfaction and reactive power usage. The ORPF is the theoretical optimum. Coordinated controllers such as OFO can in principle achieve this optimum, whereas local controllers are subject to Pareto curves that are strictly dominated by the optimum.

The Pareto front of the performance of local controllers in Fig. 8 is the clearest illustration of the intrinsic limitation: there is a persistent gap between this front and the ORPF operating point. Both the MLP-optimized droop (which lies essentially on the Pareto front) and the DNN (that spans it via the tuning of $\lambda$) seem to reach the performance limit that distributed controllers can achieve.

## 6. Experimental results on grid enhancement

The grid enhancement capabilities of droop control and OFO were also experimentally tested on a distribution grid in Roskilde, Denmark. A detailed description of the experimental setup and the OFO implementation can be found in [17]. In short, an active power source was connected at the end of a long feeder, see Fig. 9.

Its active power injections lead to a voltage increase that limits the active power that can be injected. This power source and two reactive power sources along the feeder perform volt/var control to regulate the voltage. Fig. 10 shows the measured voltage at the end of the feeder for different active power injections. The lines correspond to the different control methods used. The figure shows that without any volt/var control, the voltage limit of 1.05 p.u. is reached at 8.77 kW. Using droop control, the limit is reached at 10.16 kW. When OFO is used for control, the grid can be enhanced to transmit 11.23 kW which is another 10.5% on top of the capacity achieved using droop control.

Finally, we tested whether the ORPF problem could be solved in real time and used as a feedforward controller, that is, by directly feeding the solution of the optimization problem to the DERs as reactive power setpoints. Despite the theoretical optimality of the ORPF solution, the resulting curve (red line) violates the voltage constraint. This is due to a model mismatch that occurred even though the cable types and parameters were known and all generation and consumption were measured. This showcases that feedforward methods fail to enforce constraints in real systems even under excellent model knowledge. In contrast, OFO can track the ORPF solution and guarantees constraint satisfaction even under model mismatch, thanks to its feedback nature that uses real-time voltage measurements from the grid.

The experiment is intended as a proof-of-concept corroborating the qualitative findings of the simulation study. In particular, it demonstrates

- the potential of coordinated volt/var control compared to standard droop control
- the need for a feedback control architecture and the fragility of simplistic feedforward optimization solutions such as ORPF for the operation of distribution grids.

The feeder used is small, with only three controllable devices, and the results should not be interpreted as a demonstration of scalability to full-scale distribution networks. A real-world validation at that scale remains a direction for future work.

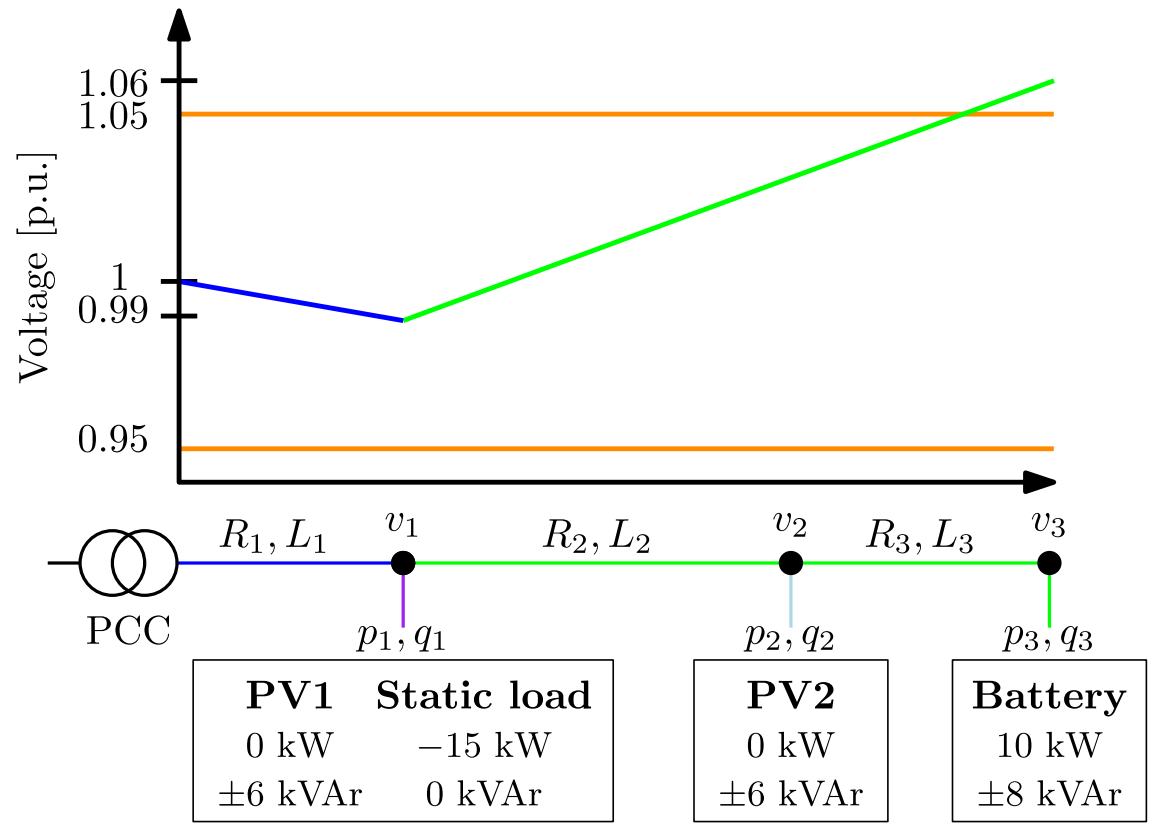

**Fig. 9.** Topology of the real distribution grid feeder used for the experimental validation. The battery acts as renewable generation, PV2 is able to provide reactive power injections, and PV1 and a static load mimic an electric vehicle. The figure is taken from [17].

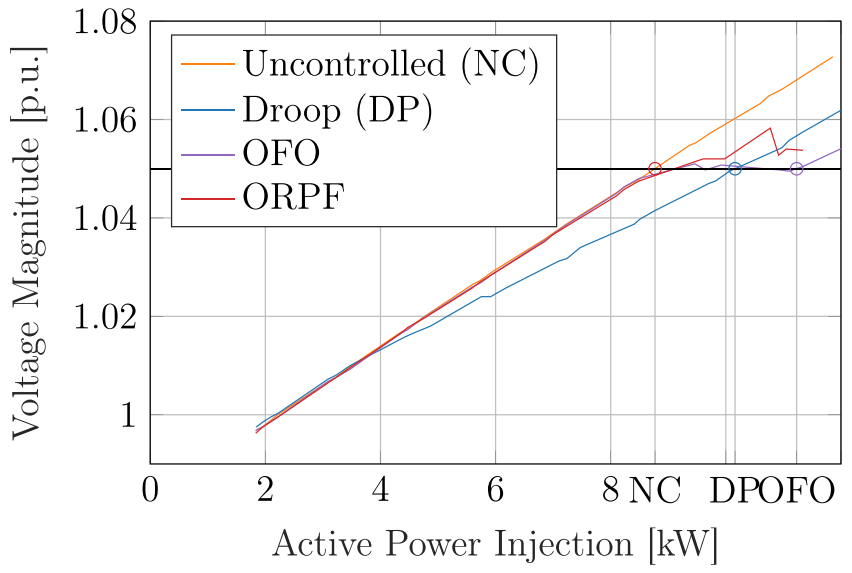

**Fig. 10.** Experimental results of grid enhancement through volt/var control on a real distribution grid feeder. The OFO controller prevents persistent constraint violations up to a total active power injection of 11.23 kW. At this point, the reactive power capabilities finally saturate and no additional reactive power can be absorbed at any bus.

## 7. Conclusion

The experiment and simulations show that volt/var control is an effective means to enhance the capacity of power distribution grids. It enables more active power to be transmitted before voltage constraints are violated and does so cost-effectively. Consequently, volt/var control will be a crucial component of distribution grids in the near future, as the installed capacity of distributed energy sources such as PV increases even further. However, our experiments demonstrate that the current state of the art, droop control, is inherently suboptimal, both with respect to the constraint violations mitigated and the efficient use of reactive power.

Advanced local control methods such as the presented optimized droop controllers have evident advantages over standard droop control. They can reduce the optimality gap without the need for a communication infrastructure. How close they get to optimality depends on the given method as well as external factors. Typically, no guarantees can be made about their performance under unexpected operating conditions. Other promising lines of research investigate local data-driven control methods beyond the tuning of droop curves [36–39]. While these were not included explicitly in this work, it remains interesting to explore by how much and under what conditions these methods can further reduce the optimality gap. We encourage future studies to replicate our experiments to validate and extend their findings.

In contrast to local control, we demonstrated how adding communication and coordinating the volt/var control across the whole grid can yield near-optimal mitigation of constraint violations using minimal reactive power. The steady-state simulations reveal that the grid capacity can be enhanced by 10.4% compared to droop control when a communication channel to all reactive power resources is available. In the experiment on a real distribution grid feeder, the enhancement was 10.5%. We showed that coordinated control methods like OFO can reach this maximum level of grid enhancement using only voltage magnitude measurements and minimal model information. By using real household electricity consumption and PV production data, we showed that OFO is robust to unmeasured disturbances and performs well in a realistic setting. Finally, we demonstrated its practical viability on a real distribution grid feeder.

Overall, our analysis suggests that advanced volt/var control can enhance the capacity of a distribution grid by approximately 10% beyond the currently established practice, enabling distribution grid operators to mitigate or postpone physical grid reinforcements that may be required in the future. We believe that our study can motivate distribution grid operators to adopt more advanced volt/var control strategies and, potentially, invest in real-time communication infrastructure to enable coordinated control.

## CRediT authorship contribution statement

**Jonas G. Matt:** Writing – review & editing, Writing – original draft, Visualization, Software, Investigation. **Lukas Ortmann:** Writing – review & editing, Supervision, Project administration, Conceptualization. **Saverio Bolognani:** Writing – review & editing, Supervision, Project administration, Conceptualization. **Florian Dörfler:** Writing – review & editing, Supervision, Project administration, Funding acquisition, Conceptualization.

## Declaration of Generative AI and AI-assisted technologies in the writing process

During the preparation of this work, the authors used AI tools for spell- and grammar-checking, as well as style improvements. The authors reviewed and edited the content as needed and take full responsibility for the content of the published article.

## Declaration of competing interest

The authors declare that they have no known competing financial interests or personal relationships that could have appeared to influence the work reported in this paper.

## Appendix A. Training on past scenarios

Both DNN- and MLP-optimized droop control rely on historical operating data for training. In the main experiments, we use half the available data (every other day) for training and evaluate performance on the full dataset. However, this data was not truly “historical”: The droop curves evaluated for a given PV integration scenario were optimized using the data from the same scenario. It is well known in machine learning that testing on training data yields an optimistic estimate of achievable performance.

To assess the robustness of these methods under more realistic conditions, we evaluate the DNN-optimized droop controller when trained on data from different PV integration scenarios. Fig. A.11 illustrates the performance when each DNN is trained exclusively on the data of one scenario (2020, 2030, or 2035), and then tested across all scenarios.

This setup mimics the practical situation where droop curves are trained once using data from a given year and subsequently deployed without regular retraining. As expected, the best performance is achieved when the training and testing scenarios match. Moreover, using past scenarios as training data generally leads to worse performance than training on future scenarios. In particular, because almost no overvoltages occur in the 2020 scenario, droop curves trained on such data are poorly adapted to the more severe overvoltage conditions in the 2030 and 2035 scenarios.

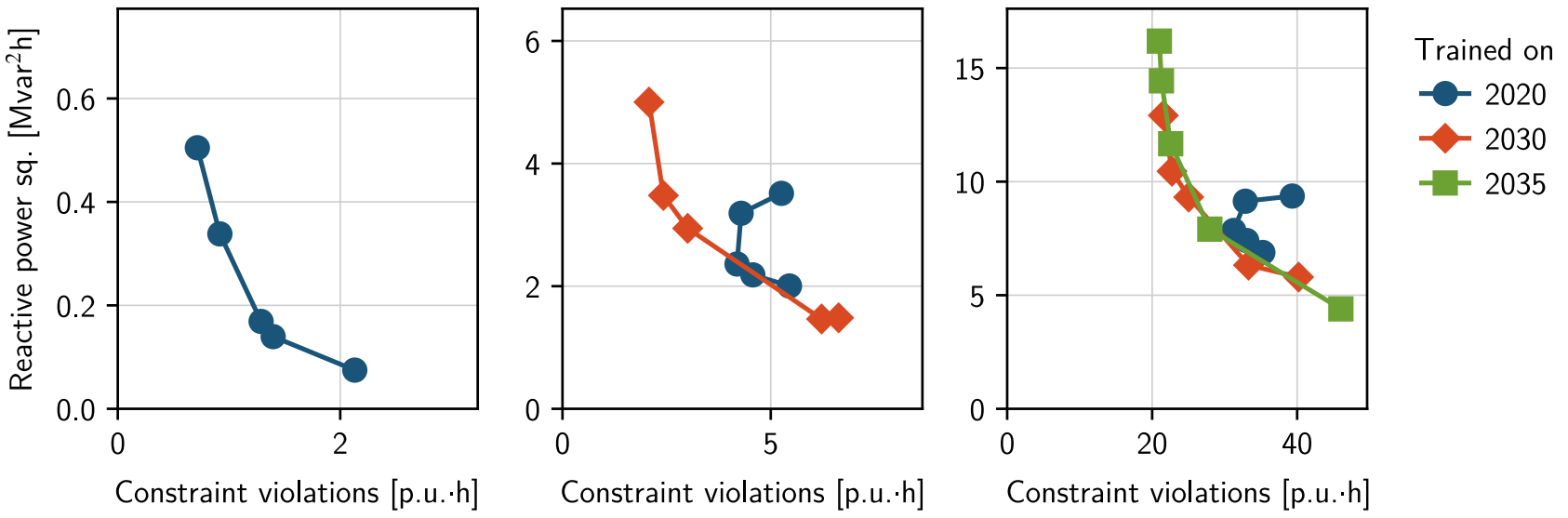

(a) Deployed in 2020 (b) Deployed in 2030 (c) Deployed in 2035

**Fig. A.11.** Performance of DNN droop control when trained on different PV integration scenarios. Each curve shows the Pareto front resulting from DNNs trained on data from one scenario and evaluated across all scenarios; each panel corresponds to evaluation on one scenario.

## Data availability

The authors do not have permission to share the data.